\documentclass[%
reprint,
amsmath,amssymb,
aps, 
superscriptaddress,
longbibliography, 
floatfix,
textcomp,mathcomp
]{revtex4-2}

\usepackage[dvipsnames]{xcolor}
\usepackage{graphicx}
\usepackage{dcolumn}
\usepackage{bm}
\usepackage{natbib}
\usepackage{makecell}
\usepackage{svg}
\usepackage{amsmath}
\usepackage{float}
\usepackage{booktabs}

\usepackage{multirow, tabularray}
\usepackage[caption=false]{subfig}
\usepackage[normalem]{ulem}
\useunder{\uline}{\ul}{}

\renewcommand{\figurename}{Fig.}
\renewcommand{\tablename}{Table}
\renewcommand{\thesection}{\arabic{section}}
\renewcommand{\thetable}{\arabic{table}}

\newcommand{\micm}{\:$\mu$m\:}

\newcommand{\ohm}{\:$\Omega$\:}
\newcommand{\siox}{SiO$_2$\:}

\newcommand{\affilSandia}[0]{\affiliation{Sandia National Laboratories, Albuquerque, New Mexico 87185, USA}}

\newcommand{\affilDuke}[0]{\affiliation{Department of Electrical and Computer Engineering, Department of Physics, Duke Quantum Center, Duke University, Durham, NC 27708, USA}}

\begin{document}
	

\title{Multi-junction surface ion trap for quantum computing} 

\author{J.D. Sterk} \email{jdsterk@sandia.gov} \affilSandia
\author{M.G. Blain} \thanks{Present address: Quantinuum, 303 South Technology Court, Broomfield, Colorado 80021, USA} \affilSandia 
\author{M. Delaney} \affilSandia
\author{R. Haltli} \affilSandia
\author{E. Heller} \affilSandia
\author{A.L. Holterhoff} \affilSandia
\author{T. Jennings} \affilSandia
\author{N. Jimenez} \affilSandia
\author{A. Kozhanov} \affilDuke
\author{Z. Meinelt} \affilSandia
\author{E. Ou} \affilSandia
\author{J. Van Der Wall} \affilSandia
\author{C. Noel} \affilDuke
\author{D. Stick} \affilSandia

\date{\today}

\begin{abstract}

Surface ion traps with two-dimensional layouts of trapping regions are natural architectures for storing large numbers of ions and supporting the connectivity needed to implement quantum algorithms. Many of the components and operations needed to fully exploit this architecture have already been demonstrated, including operation at cryogenic temperatures with low heating, low excitation transport, and ion control and detection with integrated photonics.  Here we demonstrate a trap that addresses the scaling challenge of increasing power dissipation as the RF electrode increases in size.  By raising the RF electrode and removing most of the insulating dielectric layer below it we reduce both ohmic and dielectric power dissipation.  We also measure heating rates across a range of motional frequencies and for different voltage sources in a trap with a raised RF electrode but solid dielectric.
\end{abstract}

\maketitle

\section{Introduction}

Creating more powerful trapped-ion quantum computers requires the storage, transport, and optical addressing of larger numbers of ions than can be supported by simple linear traps.  Not only does a 2D arrangement increase the number of trappable ions, but it also naturally accommodates surface codes for quantum error correction (QEC), in which qubits are arranged in a lattice with interspersed ancilla and data ions \cite{fowler:2012} or are stored in a linear array with all-to-all connectivity \cite{egan:2021}. Stabilizer circuits can be implemented through a repeated pattern of transporting ancilla ions to each of their four neighboring data ions and performing entangling operations \cite{kang:2023} (or directly applying gates gates in the linear chain), and can be tiled to form larger QEC circuits.  This topology has been developed and advanced primarily by superconducting qubit technologies \cite{google:2023}, where interactions between qubits are constrained to relatively close neighbors, if not just nearest neighbors.  While the 2D ion array can accommodate that mode of operation, ion transport can also be used for longer distance transport, such as transversal logic gates that involve moving all physical qubits within a logical qubit \cite{bluvstein:2024}.  This capability could be used to implement low density parity check (LDPC) codes \cite{bravyi:2023} that are more efficient at achieving fault tolerance with fewer qubits. 

The vision of a 2D trap array was identified early on in the ``QCCD'' architecture \cite{kielpinski:2002, metodi:2005}, and there have been many experimental demonstrations that advanced that vision since the initial conceptualization of the surface ion trap \cite{chiaverini:2005}. These include demonstrations of fundamental quantum building blocks~\cite{home:2009}, high speed linear transport \cite{sterk:2022, clark:2023}, junction shuttling \cite{burton:2023}, cryogenic operation \cite{spivey:2021}, and the integration of photonics \cite{mehta:2020,niffenegger:2020,ivory:2021,setzer:2021}.  Other demonstrations necessary for trap scaling have also occurred, including the delivery of electrical signals to islanded control and RF electrodes \cite{tabakov:2015, zhao:2021}, electrode co-wiring \cite{blain:2021, malinowski:2023}, inter-chip transport \cite{akhtar:2023}, and the integration of voltage sources \cite{stuart:2019,guise:2014}. These traps can also support modular architectures that rely on photonic interconnects \cite{monroe:2014}. Finally, there have also been demonstrations that put many of these components together to perform full algorithms \cite{moses:2023,zhu:2023,richerme:2023}. 

One of the remaining hurdles to making truly scalable surface ion trap arrays is to resolve the issue of RF power dissipation.  CMOS fabricated surface ion traps require RF voltage amplitudes typically between 80~V and 300~V at 30~MHz to 100~MHz.  The surface trap geometry results in a relatively large capacitance between the RF electrode and surrounding RF-grounded electrodes, which in turn leads to significant current and therefore ohmic power dissipation. This power dissipation scales as $P_o \sim V^2 \Omega^2 C^2 R$, where $V$ is the voltage amplitude, $\Omega$ the drive frequency, $C$ the capacitance, and $R$ the RF electrode resistance.  In addition there is power dissipation in the dielectric insulator of the trap that scales as $P_d \sim V^2 \Omega C_d \tan{\delta}$, where $\tan{\delta}$ is the loss tangent of the insulating dielectric and $C_d$ is the capacitance of just the regions that are supported by the dielectric.  If we consider a system with $n$ trapping sites, each of which requires a fixed length of RF electrode, then the total power dissipation scales as $P_{\text{total}} \sim \alpha_o n^3 + \alpha_d n$.  Here $\alpha_o$ and $\alpha_d$ are factors dependent on the RF voltage applied and the electrical properties of the RF electrode, and are associated with the ohmic and dielectric losses, respectively.

The unfavorable $n^3$ scaling can be addressed in several ways.  One way is to segment the RF electrode and use multiple RF launches, such that the number of sites per launch is held fixed and more launches are added as the number of ions grows.  In this case the ohmic power dissipation would grow linearly with the number of launches, and therefore the total number of ions. This technique is only effective if the RF distribution and lead itself do not cause significant power dissipation, and it also introduces new issues regarding phase and voltage alignment between separate RF electrodes that are adjacent. Such an approach is necessary in a quantum computer consisting of multiple modules~\cite{akhtar:2023}. Another way to reduce ohmic power dissipation is to use a substrate such as sapphire or quartz \cite{bautistaSalvador:2019, seidelin:2006} that does not require a ground plane below the RF electrode and therefore has much lower capacitance.  A drawback to this approach is that non-silicon substrates can limit the inclusion of CMOS-compatible integrated technologies. Finally, ohmic power dissipation can also be reduced using thicker metals or operating at cryogenic temperatures to lower the resistance of the RF electrode.

Here we demonstrate a large multi-junction ion trap, called the \emph{Enchilada} trap, in which ohmic and dielectric power dissipation is reduced by raising the RF electrode and removing extensive amounts of the underlying dielectric insulator, leaving only supporting pillars. This multi-junction ion trap was designed to store 200 ions in up to four multi-ion linear chains.  Heating rate measurements are performed on a version of the trap with a raised RF electrode with a solid dielectric. The scaling of the axial heating rate versus frequency is compared between experiments using different low-pass filters and voltage sources.

\section{Design and fabrication}
\label{sec:designFab}

\begin{figure*}[htbp]
    \includegraphics[width=1\textwidth]{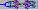}
    \caption{Layout of the \emph{Enchilada} trap, showing the entire isthmus width but only a fraction of its length.  The electrodes are color-coded according to their co-wiring.  The rectangular electrodes (green) in the center are independent, while the four side electrode sections (blue) are co-wired, as are the junctions (red).  The larger electrodes outside of the bright blue RF electrode are used for compensating electric fields and rotating the principal axis.  All gray electrodes are grounded on the trap, including the light gray background.}
    \label{fig:enchiladaLayout}
\end{figure*}

The trap uses a high-optical-access bowtie-shaped platform with a central isthmus of width 1.65 mm to accommodate lasers that skim the surface and illuminate ions 66\micm above the RF electrode. The ion height above the control electrode plane (72.1\micm\!) and isthmus half-width define a maximum side-optical access numerical aperture (NA) of 0.087, and beams with significantly smaller NA will have minimal scattering on the edges of the trap. For example, a 532~nm ideal Gaussian beam directed perpendicular to the isthmus and with a 5\micm waist at the ion will have a 0.034 NA, corresponding to a fractional power clipping on the edge of the isthmus of -67~dB below the total power. 
The device consists of six Y-junctions with five long linear sections, illustrated in Figure~\ref{fig:enchiladaLayout}. Each linear section is long enough to hold 50 ions. The outer linear sections each contain a small loading hole for loading ions in each arm independently, while the central linear section contains two holes near the two junctions.

This trap uses significant co-wiring of control electrodes in order to fit on a package with 100 electrical I/O.  There are 302 total control electrodes, with independent control of 20 side electrodes (co-wired across each side), 20 independent central electrodes, and 35 independent junction electrodes (co-wired across 6 junctions). Figure~\ref{fig:enchiladaLayout} shows how the control electrodes are co-wired.  With co-wiring, only 75 independent voltage signals are needed to control this trap.  This co-wiring does limit the regions of the trap that can be used simultaneously; for full independent operation a version with all independent control is planned.

A typical trap like this could be realized in five metal layers, with all RF and control electrodes on the top metal layer, \siox separating each metal layer, and a bottom ground plane to screen RF voltages from the lossy silicon substrate. To increase trap strength and reduce capacitance over this design, we added an additional metal layer (M6) to raise the RF electrode above the other electrodes. Here the top and bottom metal layers are designed to be 2.55\micm thick, while all other metal layers are 1.35\micm thick. Each consists of 2.4\micm or 1.2\micm aluminum (with 1/2\% copper) sandwiched between thin titanium nitride layers.  The two topmost \siox layers are 4\micm thick, while all others are 2\micm thick.  The closest ground below the RF electrode is on the third metal layer (M3); it serves to screen the control electrode leads on M2 from picking up RF signals.

\begin{figure*}[htbp]
    \includegraphics[width=1\textwidth]{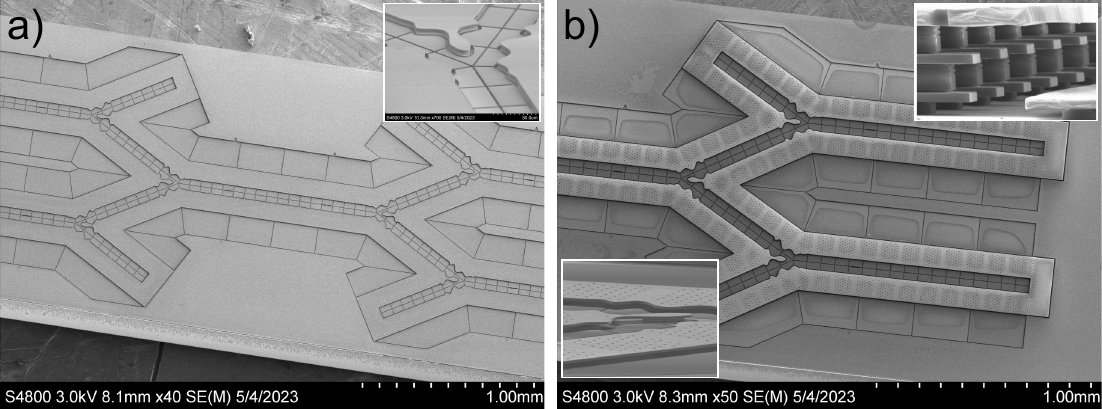}
    \caption{SEM micrograph images of a (a) solid and (b) perforated trap.  In part (a) the majority of the trap is visible, with an inset showing the raised RF electrode in the junction.  Part (b) shows the perforated traps.  The holes to allow etchant access are visible in the RF electrode, as are the pillars, notable due to the absence of holes.  The lower left inset shows how the RF electrode in the junction is suspended with a gap in the \siox\!, while the inset in the top right shows the \siox pillars in a linear region.  The pillars of oxide land on floating metal in layers M5 and M4, visible as finger-shaped structures in the inset.}
    \label{fig:enchiladaSEM}
\end{figure*}

Given the lateral geometry of this design, we would expect a 5 layer trap to have 11.9~pF of total capacitance between the RF electrode (including lead) and the ground plane in M3.  By adding an additional metal and oxide layer to raise the RF electrode 4\micm above the control electrode layer, and undercutting 5\micm of oxide from the edge of the RF electrode, the calculated capacitance drops to 6.5~pF.  This raised-RF geometry has been demonstrated in the past \cite{leibrandt:2009} as well as in a device where only the interior control electrodes were lowered \cite{revelle:2020}. The capacitance can be decreased even further by eliminating most of the \siox below the RF electrode and only leaving \siox pillars for structural support. These pillars are created using the same controlled-etchback technique employed to define the vertical \siox sidewalls \cite{blain:2021} in the solid version of the trap. To allow access for the release etchant to remove all \siox outside of the structural pillars, the top metal RF was perforated.  For the perforated design, 80\% of the oxide was removed below the trap RF electrode (but not the lead), reducing the calculated total capacitance to 3.7~pF (2.3~pF for just the trap region). Fig.~\ref{fig:enchiladaSEM} shows both solid (a) and perforated (b) versions of the trap. 

The RF electrode is 100\micm wide with a separation of 73\micm\!.  All interior control electrodes in the linear sections are 62\micm long and 30\micm wide. All lateral gaps between RF and control or ground electrodes are 5\micm\!, while all gaps between control or ground electrodes and themselves are 3\micm\!.  Simulations showed that by raising the RF electrode 4\micm above the surrounding control electrodes, the ion height moves down by 6.5\micm (from 72.2\micm to 65.7\micm above the top of the RF electrode) and the radial frequency increases by 17\%, allowing for less voltage and 34\% less power dissipation. 

The radial motional frequency for an ion of mass $m$ and charge $q$ in this trap is
$\omega=\frac{q V}{\sqrt{2}\Omega m \Lambda^2}$, where $V$ is the voltage amplitude and $\Omega/2\pi$ the frequency of the applied RF voltage. The characteristic distance $\Lambda$ for this trap is 124\micm\!; $1/\Lambda^2$ describes the local curvature of the electrostatic potential arising from the RF electrode at the pseudopotential minimum. It is given by the inverse square root of the eigenvalues of the Hessian of the RF potential per volt and is expressed this way to parameterize the reduced efficacy of the surface trap relative to a true hyperbolic geometry (in which case $\Lambda$ equals the ion--electrode distance). The depth in the radial directional equals $\frac{1}{2} \alpha m \omega^2 \Lambda^2$, where $\alpha=.023$ and reflects the reduced efficacy of the surface trap geometry. 

While our experiments used calcium, these traps were designed to support heavier ions like ytterbium and barium.  The ion species choice does not affect the electrode geometry, but it does influence design choices that impact voltage breakdown and power dissipation.  Based on the geometry of the RF electrodes and the application of 300~V RF amplitude at 50 MHz (which would generate a 4.7 MHz radial trapping frequency in the linear regions), we calculate the power dissipations displayed in Table \ref{tab:power}.  In this table, the ohmic and dielectric losses are separated, as they have different dependencies on frequency and capacitance:
\begin{eqnarray*}
P_{\text{ohmic}} & = & \frac{1}{2} V^2 \Omega^2 C^2 R/3 \\
P_{\text{dielectric}} & = & \frac{\Omega}{2} \int \tan(\delta) \varepsilon_0 \varepsilon_r |\mathbf{E}|^2 \, d^3\mathbf{x} \\
& = & \frac{1}{2} V^2 \Omega C_{\text{ox}} \, \tan(\delta),
\end{eqnarray*}
where $R$ and $C$ are the lumped element resistance and capacitance, $C_{\text{ox}}$ is the capacitance of just the regions with oxide pillars, $\varepsilon_0$ is the permittivity of free space, $\varepsilon_r$ is the relative permittivity of \siox\!, and $\tan(\delta)$ is the loss tangent of the oxide ($10^{-3}$ for the oxide used here).  The volume integral for dielectric power loss is taken over the region between the RF and the underlying ground plane, and vanishes except where there is dielectric with a non-zero $\tan(\delta)$ (i.e., just the \siox pillars).
The factor of $1/3$ in the ohmic power calculation is due to the distributed nature of the resistance and capacitance in the RF electrode.

The table calculations also separate the power loss in the trap region from the power loss in the lead, and are repeated for both a perforated trap and a solid trap. It should be noted that the perforation technique does reduce the thermal conductivity from the electrode to the substrate, and so even if the power is substantially reduced, the electrode temperature may not drop as significantly if the thermal conductivity is also reduced.

\setlength{\tabcolsep}{10pt} 
\begin{table}
\caption{Calculated power dissipation in the \emph{Enchilada} trap.  For the solid-dielectric version, there is oxide under the entire RF electrode up to 5\micm from each edge.  For the perforated-dielectric trap, oxide pillars are under 20\% of the RF electrode, with the rest vacuum.} 
\label{tab:power}
\begin{tabular}{@{} rccccc @{}} 
 \toprule[1pt] 
 \midrule[0.3pt]
 & \multicolumn{2}{c}{Solid} && \multicolumn{2}{c@{}}{Perforated}\\
 \cmidrule(lr){2-3} \cmidrule(l){5-6}
            & Trap & Lead && Trap & Lead \\
 \midrule 
 Ohmic (mW)  &  9.7 & 13.4 && 1.9 & 4.5\\
 \midrule
 Dielectric (mW)  &  59.1 & 19.4 && 12.9 & 19.4\\
 \midrule
  \textbf{Total (mW)}  & \multicolumn{2}{c}{\textbf{101.6}}  && \multicolumn{2}{c}{\textbf{38.7}}\\
 \midrule[0.3pt]
 \bottomrule[1pt] 
\end{tabular}
\end{table}

\begin{figure}[htbp]
    \includegraphics[width=.48\textwidth]{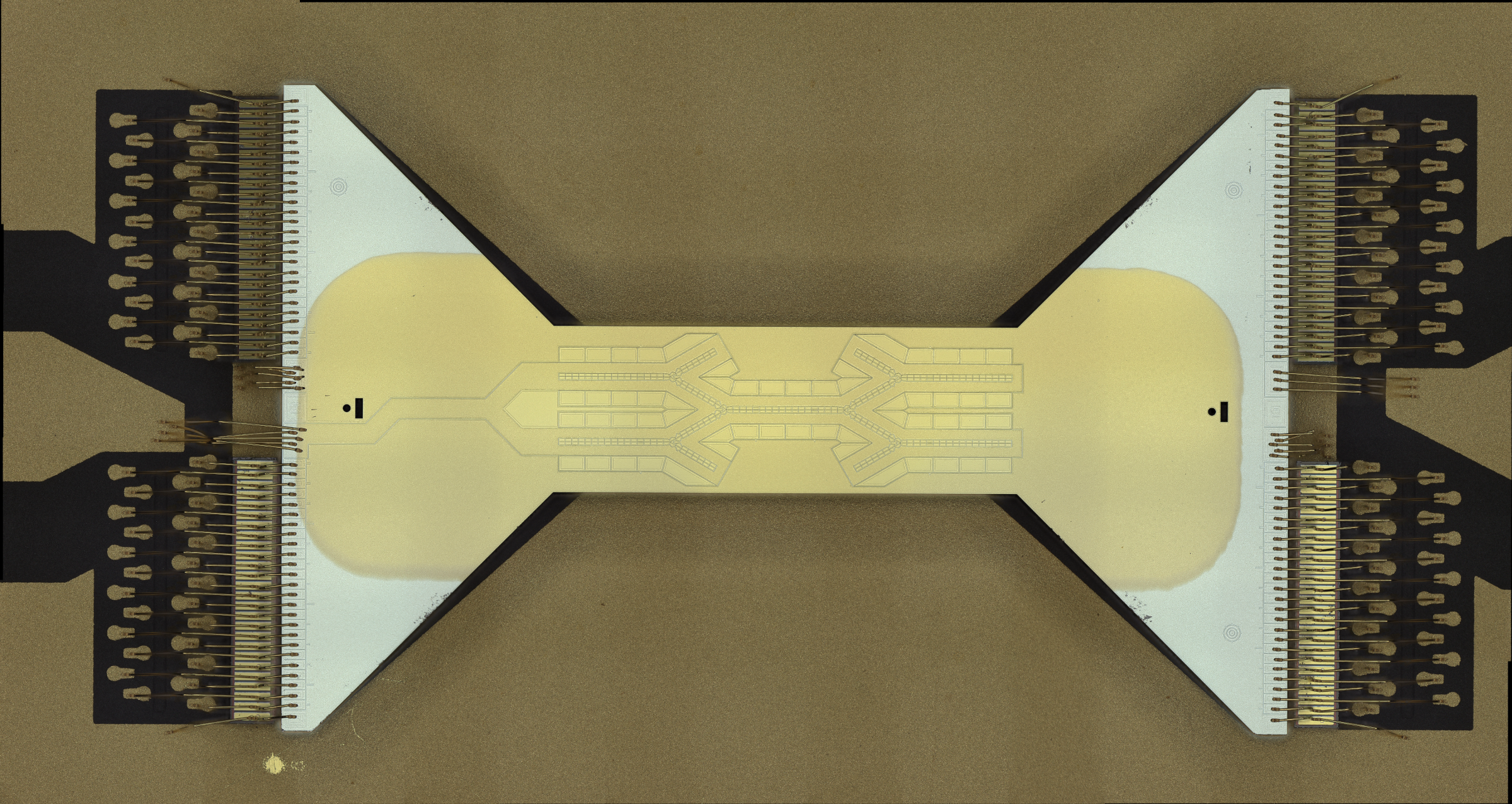}
    \caption{A packaged trap.  The central part is gold coated, with a physical mask used to protect the die bond pads from gold coating that could cause purple plague. The wirebonder connects the die bond pads to trench capacitors on the four chiplets (two at each end of the trap chip) and then the package pads.}
    \label{fig:enchiladaPackage}
\end{figure}

Each one of the control electrodes is RF grounded using trench capacitors \cite{blain:2021}. To reduce fabrication complexity, they were fabricated on four separate chiplets rather than directly on the trap, as shown in Fig.~\ref{fig:enchiladaPackage}. The chiplets are 400\micm wide by 2600\micm long by 350\micm thick, and are solder attached to the package between the trap chip and the package bond pads, on both ends of the chip.  Each of these long narrow chiplets contain 25 individual trench capacitors that are each 95\micm $\times$ 380\micm in size and have 311 pF of capacitance. They can safely handle up to $\pm$ 20 V and have a breakdown voltage of $\pm$ 30 V. Each trench capacitor is wirebonded to both a package pad and bondpad on the trap. 

\section{Motional heating}
\label{sec:experiment}

Motional heating of a trapped ion depends on the power spectral density of electric field noise at the trapping location. Specifically, for a trap whose principal axes are given by orthonormal vectors $\hat{\mathbf{x}}_{i}$ and corresponding secular frequencies $\omega_{i}$, the heating rate is 
\[
\dot{\bar{n}}_{i} = \frac{q^{2}}{4m\hbar \omega_{i}} S_{E_{i}}(\omega_{i}),
\]
where $E_{i} = \mathbf{E} \cdot \hat{\mathbf{x}}_{i}$ is the component of the electric field noise along the principal axis.
The electric field noise is an incoherent sum of various field sources such as technical noise from the control electronics, Johnson noise from resistive elements, anomalous voltage noise from the trap surface, and other sources inside the experimental apparatus \cite{brownnutt:2015}.

In our characterization experiments, we measured the axial heating rate of a single trapped calcium ion at the center-right loading hole as a function of axial frequency. Ions were trapped with $85~\text{V}$ of RF at $41.54~\text{MHz}$. Voltage solutions for a range of axial frequencies were synthesized in a manner similar to~\cite{hogle:2024} to give the desired axial frequency with a 30~degree rotation of the radial principal axes using a minimal set of electrodes. The resulting radial frequencies were $5.7~\text{MHz}$ and $6.0~\text{MHz}$.

\begin{figure}[tbp]
    \includegraphics[width=\columnwidth]{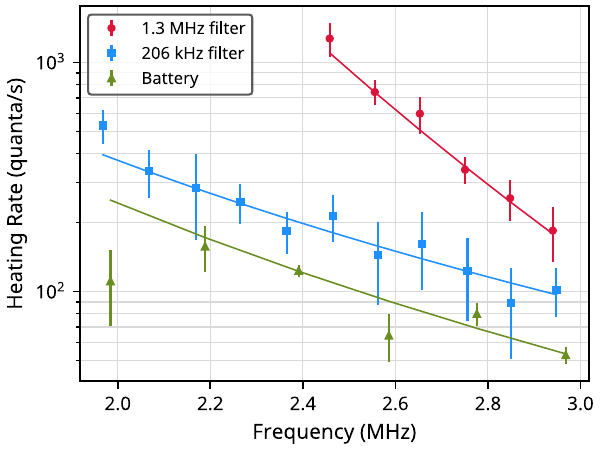}
    \caption{Heating rates as a function of axial trapping frequency. Using the same voltage solution, we measured the heating rate versus axial frequency when using a custom voltage control system with a 1.3~MHz cutoff low-pass filter (red) and a 206~kHz cutoff low-pass filter (blue).  To eliminate much of the technical noise we also measured heating when using a battery source (green) with the 206~kHz filter.}
    \label{fig:heatingrates}
\end{figure}

These measurements reveal the shape of the power spectral density $S_{E_{\text{axial}}}$ at the ion. In figure~\ref{fig:heatingrates}, we show the heating rates from $2$-$3~\text{MHz}$ for three different configurations. Heating rates were first measured using electronics designed for fast shuttling~\cite{sterk:2022} and a 6th order Chebyshev low-pass filter with a $3~\text{dB}$ cutoff-frequency of $1.3~\text{MHz}$. The power-law scaling of the heating rate for this configuration is $\dot{\bar{n}}_{\text{axial}} \propto \omega^{-10.2 \pm 1.6}$.  This steep drop in the motional heating rate with respect to frequency indicates that it is dominated by technical noise and strongly filtered by the low-pass filter, which has a 40~dB/octave roll-off.  Using the same voltage source, the heating rates were then measured with a 3rd order cascaded RC filter with a 3 dB cutoff frequency of 206~kHz and an 18~dB/octave roll-off. The data exhibited a lower heating rate due to the lower cutoff frequency, but the slower roll-off led to a lower scaling of the heating rate with frequency, $\dot{\bar{n}}_{\text{axial}} \propto \omega^{-3.5 \pm 0.7}$. Finally, the heating rate was measured with a battery and the cascaded RC filter in order to dramatically reduce the technical noise. The overall heating rate was lower but had a similar scaling as the $206~\text{kHz}$ filter data,  $\dot{\bar{n}}_{\text{axial}} \propto \omega^{-3.8 \pm 0.7}$. Due to this scaling, we conclude that the heating rate measured with the battery is dominated by voltage fluctuators on the electrode surface, as the Johnson noise limit would scale as $\omega^{-1}$.

The resistance of the battery (internal), chamber wires, package leads, and wirebonds are sub-ohm, and above 2 MHz the loockback resistance of the filter is below 2 ohms. Therefore the majority of the Johnson noise can be attributed to resistance in the leads connecting the bond pads of the trap to the electrodes. In this trap there are many electrode leads to route on M2 and therefore their width is constrained; at their narrowest they are 5\micm wide and 1.35\micm thick.  Based on the geometry of the leads (which varies), we estimate an average control electrode resistance of 12\ohm\!. The voltage noise on the $n$-th electrode arising from Johnson noise at a temperature $T$ has a power spectral density of $S_{V_{n}} = 4k_{B}TR_{n}$, where $k_{B}$ is the Boltzmann constant and $R_{n}$ is the total resistance of the lead to the $n$-th electrode. The power spectral density of the electric field at the ion is the incoherent sum~\cite{deslauriers:2006} over all $N$ electrodes, 
\begin{equation*}
    S_{E_i}(\omega) = \sum_{n=1}^{N} \epsilon_{n,i}^2 S_{V_{n}}(\omega),
\end{equation*}
where $\epsilon_{n,i}$ is a geometric factor from simulation that relates the electric field in the direction $\hat{\mathbf{x}}_{i}$ to the voltage on the $n$-th electrode. Using the estimated resistance, temperature (room), and values for $\epsilon_{n,i}$ from simulations, the calculated axial heating rate from Johnson noise is $\dot{\bar{n}}_{\text{axial}}$=15~quanta/s at $2~\text{MHz}$, below the measured heating rate.  Like the scaling, this is consistent with a dominant noise source of voltage fluctuators on the electrode surface.

\section{Conclusion}
\label{sec:conclusion}

As the ability to deliver optical control signals to larger numbers of ions grows, so does the demand for larger trap arrays. Multiple challenges arise for even slightly larger traps than are currently used, including congestion in lead routing, limitations in the number of perimeter wirebonds for electrical signal delivery, developing packages and in-vacuum sockets with sufficient electrical I/O, and power dissipation on the RF electrode.  This paper addresses the latter challenge by employing a technique to raise the RF electrode and remove insulating material under it, thereby reducing the capacitance and power dissipation on the trap. Using this technique we fabricated a large multi-junction ion trap designed to hold 200 ions and measured heating rates for various voltage sources and filters. Future research will include comparative measurements between the perforated and solid versions of the trap, as well as demonstrations of ion transport and chains of ions.

\begin{acknowledgments}
This material is based upon work supported by the U.S. Department of Energy, Office of Science, National Quantum Information Science Research Centers, Quantum Systems Accelerator.

Sandia National Laboratories is a multi-mission laboratory managed and operated by National Technology \& Engineering Solutions of Sandia, LLC (NTESS), a wholly owned subsidiary of Honeywell International Inc., for the U.S. Department of Energy’s National Nuclear Security Administration (DOE/NNSA) under contract DE-NA0003525. This written work is authored by an employee of NTESS. The employee, not NTESS, owns the right, title and interest in and to the written work and is responsible for its contents. Any subjective views or opinions that might be expressed in the written work do not necessarily represent the views of the U.S. Government. The publisher acknowledges that the U.S. Government retains a non-exclusive, paid-up, irrevocable, world-wide license to publish or reproduce the published form of this written work or allow others to do so, for U.S. Government purposes. The DOE will provide public access to results of federally sponsored research in accordance with the DOE Public Access Plan.
\end{acknowledgments}

\section*{References}
\bibliography{ion_refs}

\end{document}